\documentclass[12pt]{iopart}
\usepackage{amssymb}
\usepackage{graphicx}
\usepackage{dcolumn}
\usepackage{bm}
\usepackage[utf8]{inputenc}
\usepackage{placeins} 
\usepackage{appendix}
\usepackage[normalem]{ulem}
\usepackage[colorinlistoftodos, color=green!40, prependcaption]{todonotes}

\begin{document}
\title[Towards the Scalable Fabrication of thin-film Superconducting Paramps]{Towards the Scalable Fabrication of thin-film Superconducting Parametric Amplifiers}

\author{A.~El~Kass$^{1,2}$, K.~A.~F.~Simoes$^1$,
C.~Chua$^1$, D.~J.~Reilly$^{1,2}$, K.~Zuo$^{1,2}$ and T.~A.~Ohki$^1$}

\address{$^1$ School of Physics, The University of Sydney, Sydney, NSW 2006, Australia}
\address{$^2$ ARC Centre of Excellence for Engineered Quantum Systems, School of Physics, The University of Sydney, Sydney, NSW 2006, Australia}

\date{\today}

\begin{abstract} 

Kinetic inductance travelling-wave parametric amplifiers (KTWPAs) are emerging as core components in many applications where wideband cryogenic rf amplification at or near the quantum limit of added noise is critical. These thin film superconducting devices are unique in their ability to simultaneously provide large dynamic range and quantum-limited amplification of single photon to 100,000s of photon signals. Despite the promising performance of co-planar NbTiN thin-film KTWPAs, the original promise of a "simple" single-layer metal fabrication has encountered roadblocks and their broader adoption has been hindered by low fabrication yield. In this work, we present a post-lithography correction technique that eliminates short circuits, significantly improving yield and enabling reliable wafer-scale production. Using automated image acquisition, error analysis, and correction, we successfully fabricated operational KTWPAs on high-resistivity silicon, achieving $>$ 10~dB between 2--4 GHz. This approach paves the way for the scaling up manufacturing of KTWPAs, positioning them for widespread deployment in quantum technologies. 


\end{abstract}

\maketitle

\section{Introduction}

Superconducting quantum noise-limited amplifiers are critical components in quantum computing platforms, enabling high-fidelity readout by adding minimal amounts of noise to preserve signal-to-noise ratios essential for single-shot measurements~\cite{Aumentado2020, Krantz2019, Stehlik2015, Schaal2020, Oakes2023}. As quantum systems scale, the demand for amplifiers with wider bandwidths and higher saturation power increases to support fast multiplexed readout, a regime where KTWPAs offer significant advantages~\cite{Esposito2021, HoEom2012, Vissers2016, Chaudhuri2017, Malnou2021, Ahrens2024, Ohki2018} over Josephson junction based amplifiers such as JTWPAs~\cite{Esposito2021}. For example KTWPAs, in comparison, demonstrate near quantum-limited noise performance at high magnetic field and elevated temperature~\cite{Malnou2022, Xu2023, Frasca2024}, providing versatility in hardware integration across diverse quantum computing readout architectures. This adaptability extends to various qubit technologies, including superconducting and spin qubits, allowing operation under disparate environmental conditions such as temperatures and static magnetic fields.

KTWPAs exploit the kinetic inductance of superconducting thin films for wide-bandwidth frequency mixing and amplification. This kinetic inductance critically depends on both film width and length, requiring precise dimensional control during fabrication and highlighting the delicate nature of KTWPA production. Among KTWPA designs, co-planar waveguide (CPW) geometries stand out for their relatively efficient fabrication process. Achieving high fabrication yields for CPW-KTWPAs however remains a significant challenge. The design typically requires consistent 1~$\mu$m gaps along a 10~m "fishbone" perimeter and a transmission line spanning tens of centimeters. Even minor artefacts~\cite{Faverzani2024}  across these extended dimensions can compromise device performance. While increasing kinetic inductance in highly disordered thin-film superconductors relaxes length constraints for the transmission line, it necessitates either narrower gaps or longer interdigitated fingers to maintain impedance matching, further complicating fabrication and reducing yield.

To address these challenges, we propose a post-fabrication correction technique designed to rectify common faults encountered during CPW-KTWPA fabrication. This approach involves localized and targeted correction of fabrication errors, augmented by image processing techniques to identify and address specific defects.

Our correction solution offers several advantages over alternative methods such as the double lithography\,/\,dry etch process, a process where the design is resist-patterned and etched twice with identical or slightly different layouts. The double lithography\,/\,dry etch process, although effective in mitigating certain issues associated with random artefacts in the resist mask, was insufficient to fully resolve the fabrication challenge. Moreover, it complicates the trench morphology, significantly altering the capacitance of the spine structure and, consequently, the overall rf performance of the KTWPA. We also explored the use of a hard mask for patterning the KTWPA structure, where the superconducting film is etched through a hard mask fabricated in metal or oxide and independently prepared for this purpose, but this approach merely shifts the fabrication challenge to a different platform, increases cost, and is constrained by the initial mask design. Additionally, it does not prevent short circuits from film defects or irregularities during dry etch.

By implementing our proposed localized and targeted post-fabrication correction method, we aim to significantly improve device yield and ensure greater consistency in performance. This enhancement in fabrication reliability is expected to advance the readiness of CPW-KTWPAs for wafer-scale manufacturing for quantum computing applications and beyond~\cite{Smith2013, Pagano2022}. As quantum systems continue to progress in scale and complexity, applying these concepts more broadly towards advancing wafer-scale production of quantum devices, such as superconducting qubit processors and single photon detectors, can provide a path for higher yield manufacturing. This work demonstrates a proof concept of these techniques applied to the fabrication challenges of thin-film kinetic inductance amplifiers.

\begin{figure*}[!h]
\centering{\includegraphics[width=1\textwidth]{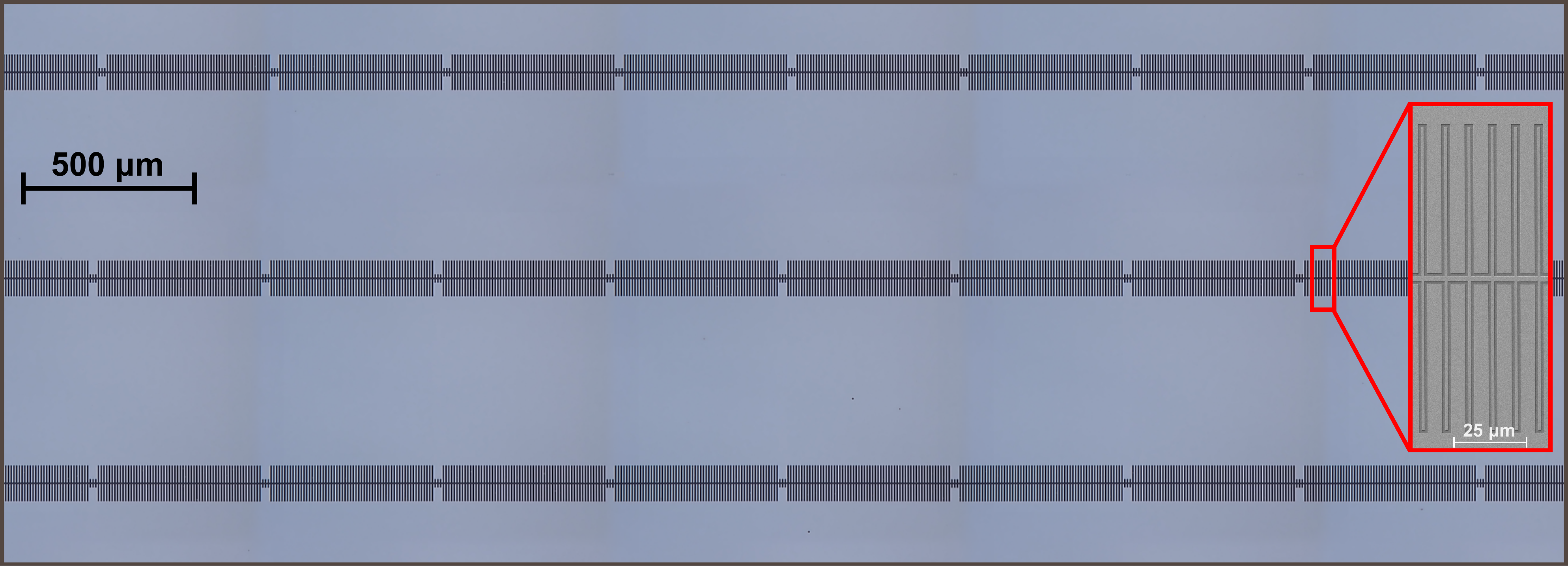}} 
\caption{A microscope photo of a section of the KTWPA chip showing a portion of 3 rows of the CPW design. The periodically repeating structure is the supercell. The inset shows a close-up SEM image of the CPW and the spine-like matching structures of the KTWPA.}
\label{fig:KTsch}
\end{figure*}

\section{Methods} \label{sec:methods}

We fabricate our KTWPA devices on a 20~nm thick NbTiN film grown via physical vapour deposition on high-resistivity Si substrate, which we have measured to have a sheet inductance of about 10~pH/$\square$. The pattern is defined using electron-beam lithography on a positive resist (ZEP520) and dry etched. Our KTWPA design consists of approximately 400 supercells spanning a total transmission line length of about 20~cm and a targeted bandwidth of 2--4~GHz.

\section{Types of Fabrication Errors}

\begin{figure*}[!h]
\centering
\includegraphics[width=1.0\textwidth]{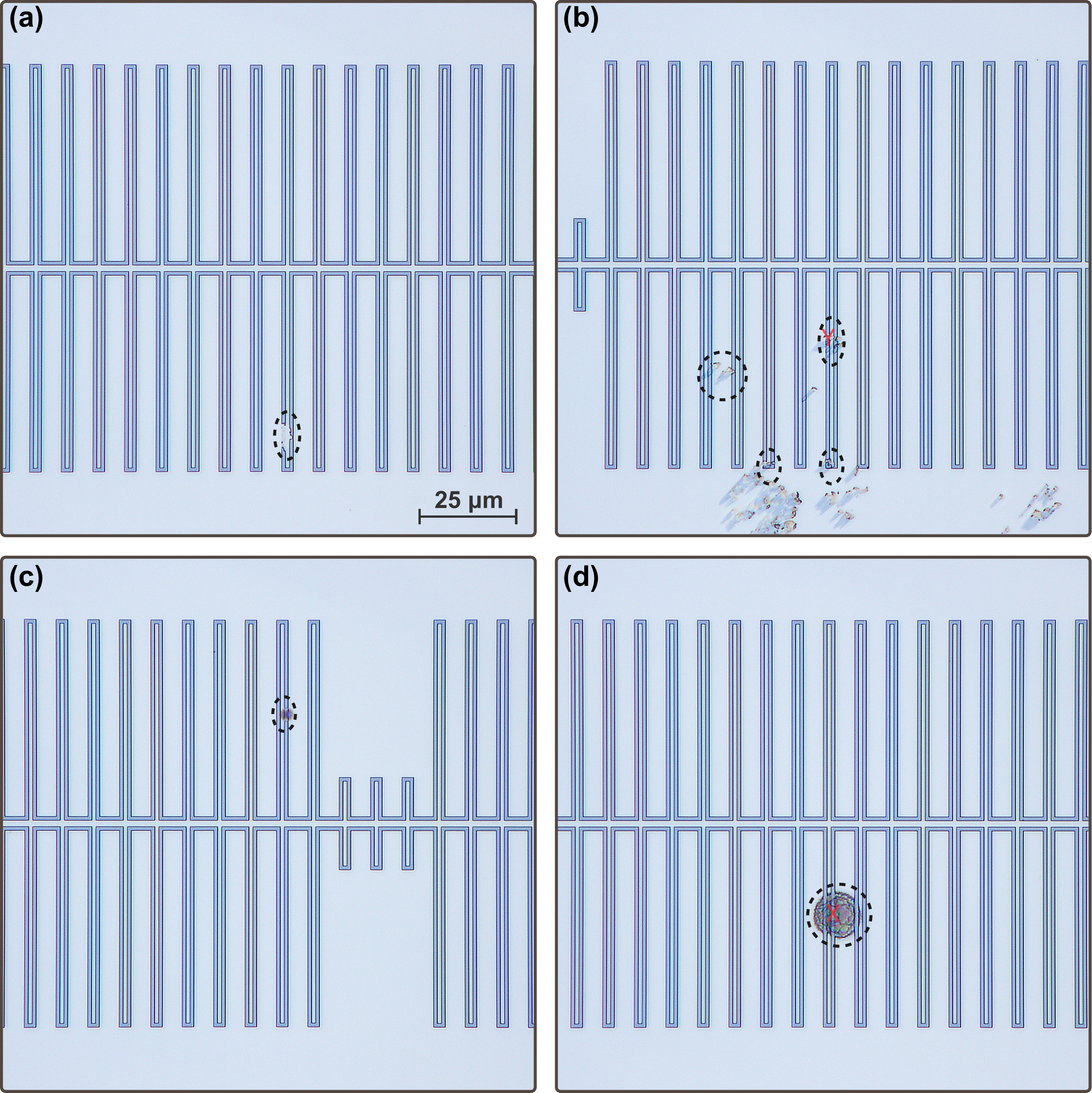} 
\caption{A number of common fabrication artefacts are shown. The faults are identified by dashed shapes manually drawn on the images to help the reader. (a) An etch fail due to an anomaly in the dry etch chamber, (b) A thick overgrowth in the film, (c) An irregular spot in the film, and (d) a resist-mask artefact. The EDS results correspond to the images (b) and (c) where Nb is detected in the gaps between the interdigitated structures with a percentage by mass $> 60\%$ in position Y and $> 20\%$ in position X.}
\label{fig:ErrorList}
\end{figure*}

A selection of recurring and potentially serious fabrication defects is shown in Fig.~\ref{fig:ErrorList}. In Fig.~\ref{fig:ErrorList}.a, the dry etch failed to isolate a finger from the ground plane, most likely due to a particle physically shielding the designated area from the plasma. This problem originates within the etcher and cannot be predicted by prior inspection of the mask. In some cases, an etch failure may also occur due to an irregularity in the mask development, resulting in a similar outcome as in Fig.~\ref{fig:ErrorList}.a, a fin short-circuited to the ground trace. The defect in Fig.~\ref{fig:ErrorList}.b was identified using localized energy-dispersive x-ray spectroscopy (EDS) in the SEM as an overgrowth in the NbTiN film. At position 'Y' in Fig.~\ref{fig:ErrorList}.b, the mass composition of Nb is found to be $>~60\%$, a proportion significantly higher than typically observed in that given film. In the particular case of Fig.~\ref{fig:ErrorList}.b, the EDS indicates that elevated levels of Nb are present in the gaps of the ``fishbone" design. Standard dry etching recipes used to process the KTWPA sample were tested and proven insufficient to eliminate the overgrown anomalies in and of themselves. Fig.~\ref{fig:ErrorList}.c shows an oxidized spot in the NbTiN film, confirmed through EDS. While largely harmless to device operation as it appears in Fig.~\ref{fig:ErrorList}.c, such oxidation originates during growth and could potentially limit the critical current of the device if present on or near the center line by creating a choke point. In extreme cases, it may render its center line entirely open circuit, both scenarios being detrimental to the overall operation of the KTWPA. Finally, Fig.~\ref{fig:ErrorList}.d shows the signature of an artefact in the resist mask during the patterning step. This speck, visible during resist mask development, effectively shields the underlying area from wet and dry etching, protecting the NbTiN film beneath. Processing the sample through the wet and dry etches of the mask and the film respectively, leaves the permanent mark seen in Fig.~\ref{fig:ErrorList}.d. An EDS analysis at the position `X' in the designed gap reveals the presence of Nb with a mass percentage $>~20\%$, similar to levels seen in the bulk areas of the film used for the KTWPA fabrication, indicating the presence of a short-circuit. 

\section{Localized Error Correction}

\begin{figure*}[!h]
\centering{\includegraphics[width=170mm]{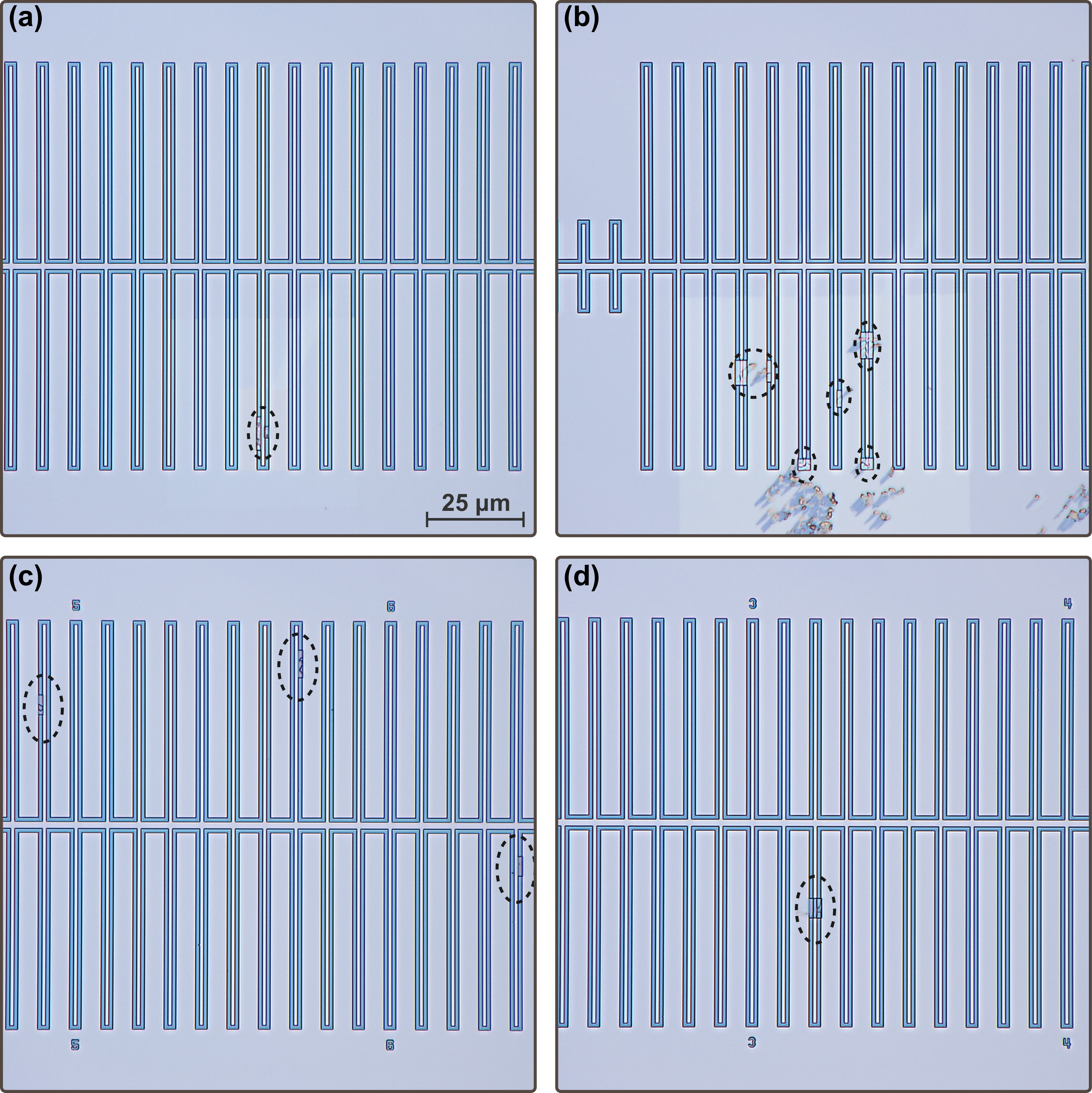}}
\caption{Localized post-lithography error correction. The corrected artefacts are identified by dashed shapes manually drawn on the images to help the reader. Panels (a) and (b) correspond to the defects in Fig~\ref{fig:ErrorList}.a and b corrected using the FIB gun in the SEM. Corrections in panels (c) and (d) were done on a different sample eliminating defects via an extra lithography\,/\,dry etch step.}
\label{fig:Correction}
\end{figure*}

The methodology for reliably fabricating functional and short circuit-free samples of the KTWPA is achieved through a post-lithography localized correction step following the main NbTiN dry etch fabrication process. This correction aims at removing the defects that may be present in the KTWPA device, such as those discussed in Fig.~\ref{fig:ErrorList}, by individually targeting specific areas of interest without causing much damage or discontinuity in the KTWPA structure, thus maintaining its engineered characteristics. After imaging and analyzing the line, a designated set of corrections is marked down for processing. Given the localized nature of the solution, the correction step could be performed either through a separate lithography\,/\,dry etch step or using focused ion beam (FIB) for physical removal of the film. Fig.~\ref{fig:Correction}.a and b show FIB-based corrections of the defects of Fig.~\ref{fig:ErrorList}.a and b respectively. Fig.~\ref{fig:Correction}.c and d demonstrate lithography-based corrections of (previously etch-failed) spots using a dry etch step.
Defect correction often involves severing the connection to the associated fin or section of the spine altogether. In some cases, it may be sufficient to remove only the artefact if possible, as seen in Fig.~\ref{fig:Correction}.a and c. For defects formed by overgrowth patches, as in Fig.~\ref{fig:Correction}.b, the artefacts themselves cannot be removed with dry etching alone. In these instances, the solution is to either use FIB to target and remove the artefacts or severe the section of the fins in question using etching or FIB alike. It is important to note that producing a high-quality KTWPA sample starts with the first step and dry etch. Extraordinary hygiene, and nearly flawless resist masks are required to obtain workable samples for correction at this phase of the fabrication, where specific human intervention is required for the analysis and decision-making regarding KTWPA structure corrections. Throughout this study, we found approximately 20--30 defects requiring correction in a typical KTWPA sample. Processing was performed on a sample-by-sample basis for the presented work. However, we expect that using established wafer-scale techniques for resist-coating and development via an automated track will improve the quality of the fabricated amplifier samples which in turn reduces the extent of the required post-processing.

\begin{figure*}[!h]
\centering{\includegraphics[width=170mm]{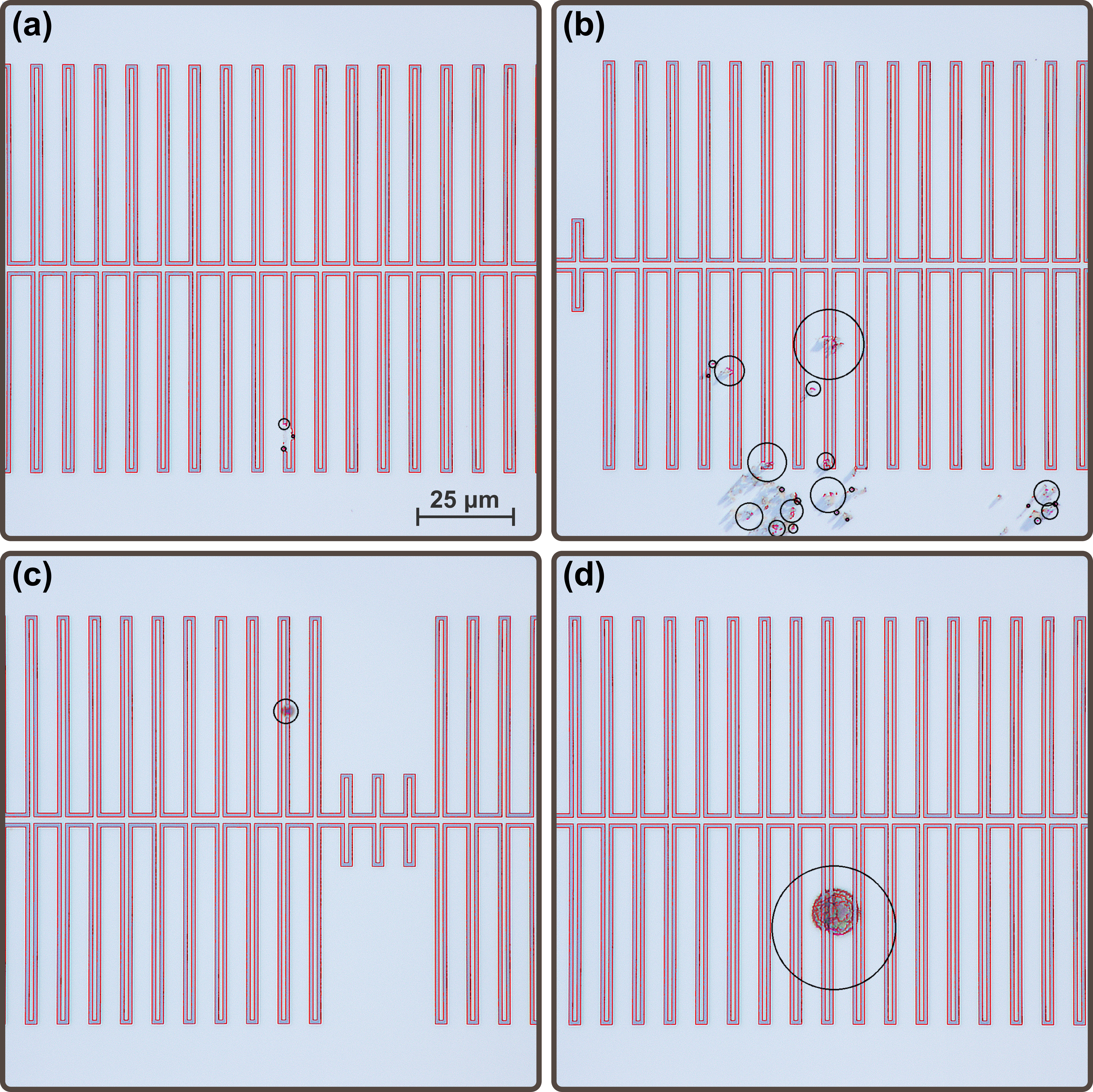}}
\caption{Image processing outcome showing edge and defect detection of the errors displayed in Fig.~\ref{fig:ErrorList}. The edges are traced in red and the defect detection outcome highlighted in dark circles, both achieved through the optimized processing described in the main text.}
\label{fig:DefDetect}
\end{figure*}

\section{Image Processing} \label{sec:img_proc}

Although specific human intervention remains predominant in the fabrication of the KTWPA at this stage, we make use of image processing techniques to assist in the analysis and correction process. Image acquisition can be automated using built-in tools within the digital microscope, allowing for the specification of image capture coordinates with the option of performing auto-focus at each position. Edge detection, a well-known technique, is applied to trace the edges of the KTWPA structure, assisting the fabrication by easily tracing the edges and providing a comparison tool for rapid error detection. The results of this processing are shown in Fig.~\ref{fig:DefDetect}, where we trace the edges and overlay them on the panels of Fig.~\ref{fig:ErrorList}. The edge detection process involves a series of post-processing steps, beginning with image segmentation. This is achieved using a k-means clustering algorithm, enabling the creation of binary image masks. By exploiting the fact that the pixel values in our raw image tend to be the darkest at the edges between etched and metal regions, we can generate a mask that closely approximates the edge sections of the image. The raw image is first converted to the Lab colour space to enhance colour differentiation before applying k-means clustering.  This alone creates a less-than-perfect mask for the edges due to noise in the image and slight fluctuations in the brightness in the edge regions. To address this, morphological operations such as closing are applied to the mask, eliminating small holes and connect disjointed edge segments caused by noise and brightness fluctuations. After refining the mask, we utilize OpenCV’s \textit{findContours} function  to detect the contours of the edges, which are then overlaid onto the original image using \textit{drawContours}, effectively highlighting the detected edges \cite{OpenCV}. Furthermore, we demonstrate specialized techniques to detect and highlight defects as indicated by the dark circles in Fig.~\ref{fig:DefDetect} generated through our code. Defect detection is accomplished by leveraging the symmetry and regular spacing inherent in the geometry of the devices. Any segment of the detected edges which does not conform to the expected regular spacing of the interdigitated fingers is presumed to be a defect. When an irregular line segment is detected, a circle is drawn about its centre point to highlight it for further inspection. If multiple segments are found in close proximity, they are clustered, and a larger encompassing circle is drawn. The defect detection process begins by identifying line segments belonging to edges in the image. The binary mask representing the edges from the previous step is fed into the Line Segment Detector (LSD) function from the OpenCV library \cite{OpenCV}, with parameters optimized for the resulting edge detection mask. Detected line segments are then clustered using the KMeans function from the scikit-learn library \cite{scikit} to find the coordinates of regularly spaced segments. Segments outside the expected coordinates corresponding to the evenly spaced lines are labelled as a defect.

\begin{figure*}[!h]
\centering{\includegraphics[width=170mm]{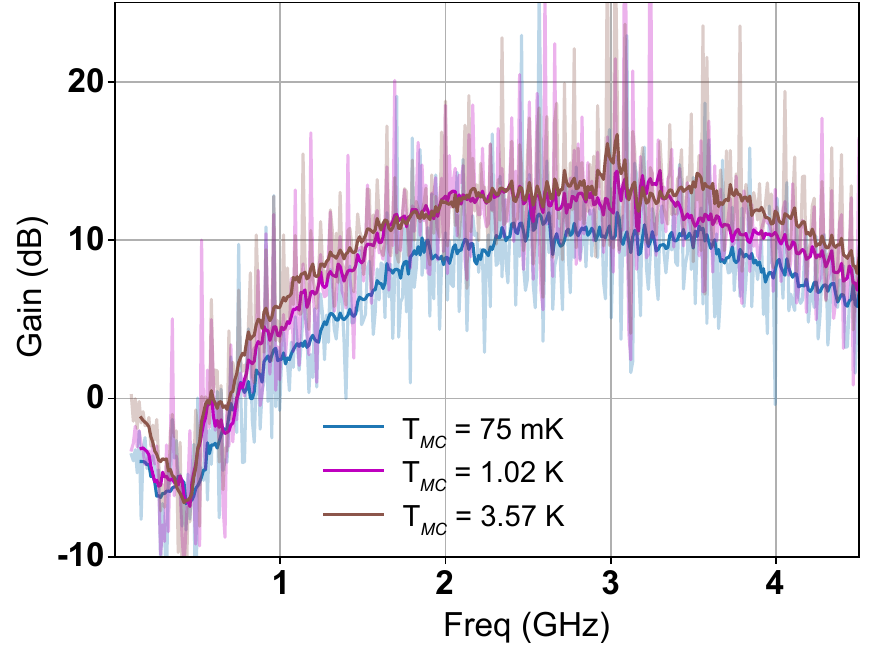}}
\caption{Gain of a successful corrected KTWPA device at 3 different temperatures in a dilution fridge.}
\label{fig:Gain}
\end{figure*}

\section{Gain Performance}

Finally, the rf behaviour of the KTWPA device is measured at cryogenic temperatures. The successfully post-processed sample is prepared for packaging with a thick TiAu ground plane deposition to reduce the effect of the thin-film ground plane on the rf performance of the device. The device is then packaged, placing as many ground plane wire bonds and stitching bonds between the turns for final testing. Initial testing verifies that the device is short-circuit free at temperatures below the critical transition of the superconducting thin film (11.5~K for this prototype). The critical current of the device is measured to be 3.8~mA at a temperature of 3.57~K and 4.0~mA around dilution fridge base temperature range (20~mK). A passive unfiltered measurement of the KTWPA transmission characteristics (S21) indicates a dispersion feature at around 5.8~GHz. The gain characteristics of the device are probed by sweeping the pump tone power and frequency range around this dispersion feature at a bias current of order 3~mA. The device demonstrates gain above 10~dB and a bandwidth between 2--4~GHz at 75~mK. Slightly higher gain is obtained at 1.02~K and 3.57~K respectively, which is an expected outcome due to the increased kinetic inductance non-linearity at elevated temperatures, as shown in Fig.~\ref{fig:Gain}. 

\section{Conclusion}

We have demonstrated a working methodology for reliable fabrication of a CPW-KTWPA sample. This approach utilizes post-lithography corrections assisted with image processing techniques to eliminate potential short circuits in the structure and ensure a defect-free continuity of the superconductor line over a perimeter with length-to-width aspect ratio of order $10^7$. The fabricated prototype shows the characteristic KTWPA signature wideband gain profile and is operable at temperatures of 4~K and below. We believe that the proposed method establishes a foundation for a short circuit-free, high-yield fabrication of KTWPA devices. This advancement has the potential to facilitate widespread adoption and use of such devices in cryogenic applications where minimal-noise high-power amplification is required.

\section{Author Attribution}
Abdallah El Kass designed the KTWPA layout, performed the full fabrication of the samples including main etch step, imaging and inspection of defects, scripting and executing the post-lithography correction pattern step, and dicing and packaging of the sample printed circuit board in copper housing. 
Abdallah El Kass, Kun Zuo and Thomas Ohki conceived the experiments, analyzed the data, and wrote the manuscript with input and assistance from Cassandra Chua, Kevin Simoes, and David Reilly.
Kevin Simoes scripted and executed the image processing.

\section{Acknowledgements}
This research was supported by the Microsoft Corporation (CPD 1-4) and the Australian Research Council Centre of Excellence for Engineered Quantum Systems (EQUS, CE170100009). The authors acknowledge the facilities as well as the scientific and technical assistance of the Research \& Prototype Foundry Core Research Facility at the University of Sydney, part of the NSW node of the NCRIS-enabled Australian National Fabrication Facility.

$\dagger$ Corresponding author: abdallah.elkass@sydney.edu.au 
\bibliographystyle{unsrt.bst}
\bibliography{AEK_cpw-twpa_paper}

\begin{thebibliography}{10}

\bibitem{Aumentado2020}
Jose Aumentado.
\newblock Superconducting parametric amplifiers: The state of the art in josephson parametric amplifiers.
\newblock {\em IEEE Microwave Magazine}, 21(8):45--59, August 2020.

\bibitem{Krantz2019}
P.~Krantz, M.~Kjaergaard, F.~Yan, T.~P. Orlando, S.~Gustavsson, and W.~D. Oliver.
\newblock A quantum engineer’s guide to superconducting qubits.
\newblock {\em Applied Physics Reviews}, 6(2), June 2019.

\bibitem{Stehlik2015}
J.~Stehlik, Y.-Y. Liu, C.~M. Quintana, C.~Eichler, T.~R. Hartke, and J.~R. Petta.
\newblock Fast charge sensing of a cavity-coupled double quantum dot using a josephson parametric amplifier.
\newblock {\em Physical Review Applied}, 4(1):014018, July 2015.

\bibitem{Schaal2020}
S.~Schaal, I.~Ahmed, J.~A. Haigh, L.~Hutin, B.~Bertrand, S.~Barraud, M.~Vinet, C.-M. Lee, N.~Stelmashenko, J.~W.~A. Robinson, J.~Y. Qiu, S.~Hacohen-Gourgy, I.~Siddiqi, M.~F. Gonzalez-Zalba, and J.~J.~L. Morton.
\newblock Fast gate-based readout of silicon quantum dots using josephson parametric amplification.
\newblock {\em Physical Review Letters}, 124(6):067701, February 2020.

\bibitem{Oakes2023}
G.~A. Oakes, V.~N. Ciriano-Tejel, D.~F. Wise, M.~A. Fogarty, T.~Lundberg, C.~Lainé, S.~Schaal, F.~Martins, D.~J. Ibberson, L.~Hutin, B.~Bertrand, N.~Stelmashenko, J.~W.~A. Robinson, L.~Ibberson, A.~Hashim, I.~Siddiqi, A.~Lee, M.~Vinet, C.~G. Smith, J.~J.~L. Morton, and M.~F. Gonzalez-Zalba.
\newblock Fast high-fidelity single-shot readout of spins in silicon using a single-electron box.
\newblock {\em Physical Review X}, 13(1):011023, 2 2023.

\bibitem{Esposito2021}
Martina Esposito, Arpit Ranadive, Luca Planat, and Nicolas Roch.
\newblock Perspective on traveling wave microwave parametric amplifiers.
\newblock {\em Applied Physics Letters}, 119(12), September 2021.

\bibitem{HoEom2012}
Byeong Ho~Eom, Peter~K. Day, Henry~G. LeDuc, and Jonas Zmuidzinas.
\newblock A wideband, low-noise superconducting amplifier with high dynamic range.
\newblock {\em Nature Physics}, 8(8):623--627, July 2012.

\bibitem{Vissers2016}
M.~R. Vissers, R.~P. Erickson, H.-S. Ku, Leila Vale, Xian Wu, G.~C. Hilton, and D.~P. Pappas.
\newblock Low-noise kinetic inductance traveling-wave amplifier using three-wave mixing.
\newblock {\em Applied Physics Letters}, 108(1), January 2016.

\bibitem{Chaudhuri2017}
S.~Chaudhuri, D.~Li, K.~D. Irwin, C.~Bockstiegel, J.~Hubmayr, J.~N. Ullom, M.~R. Vissers, and J.~Gao.
\newblock Broadband parametric amplifiers based on nonlinear kinetic inductance artificial transmission lines.
\newblock {\em Applied Physics Letters}, 110(15), April 2017.

\bibitem{Malnou2021}
M.~Malnou, M.R. Vissers, J.D. Wheeler, J.~Aumentado, J.~Hubmayr, J.N. Ullom, and J.~Gao.
\newblock Three-wave mixing kinetic inductance traveling-wave amplifier with near-quantum-limited noise performance.
\newblock {\em PRX Quantum}, 2(1):010302, January 2021.

\bibitem{Ahrens2024}
F.~Ahrens, E.~Ferri, G.~Avallone, C.~Barone, M.~Borghesi, L.~Callegaro, G.~Carapella, A.~P. Caricato, I.~Carusotto, A.~Cian, A.~D’Elia, D.~Di Gioacchino, E.~Enrico, P.~Falferi, L.~Fasolo, M.~Faverzani, G.~Filatrella, C.~Gatti, A.~Giachero, D.~Giubertoni, V.~Granata, C.~Guarcello, D.~Labranca, A.~Leo, C.~Ligi, G.~Maccarrone, F.~Mantegazzini, B.~Margesin, G.~Maruccio, R.~Mezzena, A.~G. Monteduro, R.~Moretti, A.~Nucciotti, L.~Oberto, L.~Origo, S.~Pagano, A.~S. Piedjou, L.~Piersanti, A.~Rettaroli, S.~Rizzato, S.~Tocci, A.~Vinante, and M.~Zannoni.
\newblock Development of ki-twpas for the dartwars project.
\newblock {\em IEEE Transactions on Applied Superconductivity}, 34(3):1--5, May 2024.

\bibitem{Ohki2018}
L.~Ranzani, M.~Bal, Kin~Chung Fong, G.~Ribeill, X.~Wu, J.~Long, H.-S. Ku, R.~P. Erickson, D.~Pappas, and T.~A. Ohki.
\newblock {Kinetic inductance traveling-wave amplifiers for multiplexed qubit readout}.
\newblock {\em Applied Physics Letters}, 113(24):242602, 12 2018.

\bibitem{Malnou2022}
M.~Malnou, J.~Aumentado, M.R. Vissers, J.D. Wheeler, J.~Hubmayr, J.N. Ullom, and J.~Gao.
\newblock Performance of a kinetic inductance traveling-wave parametric amplifier at 4 kelvin: Toward an alternative to semiconductor amplifiers.
\newblock {\em Physical Review Applied}, 17(4):044009, April 2022.

\bibitem{Xu2023}
Mingrui Xu, Risheng Cheng, Yufeng Wu, Gangqiang Liu, and Hong~X. Tang.
\newblock Magnetic field-resilient quantum-limited parametric amplifier.
\newblock {\em PRX Quantum}, 4(1):010322, February 2023.

\bibitem{Frasca2024}
S.~Frasca, C.~Roy, G.~Beaulieu, and P.~Scarlino.
\newblock Three-wave-mixing quantum-limited kinetic inductance parametric amplifier operating at 6 t near 1 k.
\newblock {\em Physical Review Applied}, 21(2):024011, February 2024.

\bibitem{Faverzani2024}
M.~Faverzani, P.~Campana, R.~Carobene, M.~Gobbo, F.~Ahrens, G.~Avallone, C.~Barone, M.~Borghesi, S.~Capelli, G.~Carapella, A.~P. Caricato, L.~Callegaro, I.~Carusotto, A.~Celotto, A.~Cian, A.~D’Elia, D.~Di~Gioacchino, E.~Enrico, P.~Falferi, L.~Fasolo, E.~Ferri, G.~Filatrella, C.~Gatti, D.~Giubertoni, V.~Granata, C.~Guarcello, A.~Irace, D.~Labranca, A.~Leo, C.~Ligi, G.~Maccarrone, F.~Mantegazzini, B.~Margesin, G.~Maruccio, R.~Mezzena, A.~G. Monteduro, R.~Moretti, A.~Nucciotti, L.~Oberto, L.~Origo, S.~Pagano, A.~S. Piedjou~Komnang, L.~Piersanti, A.~Rettaroli, S.~Rizzato, S.~Tocci, A.~Vinante, M.~Zannoni, and A.~Giachero.
\newblock Broadband parametric amplification in dartwars.
\newblock {\em Journal of Low Temperature Physics}, 216(1–2):156--164, May 2024.

\bibitem{Smith2013}
David~M.P. Smith, Laurens Bakker, Roel~H. Witvers, Bert~E.M. Woestenburg, and Keith~D. Palmer.
\newblock Low noise amplifier for radio astronomy.
\newblock {\em International Journal of Microwave and Wireless Technologies}, 5(4):453--461, January 2013.

\bibitem{Pagano2022}
S.~Pagano, C.~Barone, M.~Borghesi, W.~Chung, G.~Carapella, A.~P. Caricato, I.~Carusotto, A.~Cian, D.~Di Gioacchino, E.~Enrico, P.~Falferi, L.~Fasolo, M.~Faverzani, E.~Ferri, G.~Filatrella, C.~Gatti, A.~Giachero, D.~Giubertoni, A.~Greco, C.~Kutlu, A.~Leo, C.~Ligi, G.~Maccarrone, B.~Margesin, G.~Maruccio, A.~Matlashov, C.~Mauro, R.~Mezzena, A.~G. Monteduro, A.~Nucciotti, L.~Oberto, V.~Pierro, L.~Piersanti, M.~Rajteri, A.~Rettaroli, S.~Rizzato, Y.~K. Semertzidis, S.~Uchaikin, and A.~Vinante.
\newblock Development of quantum limited superconducting amplifiers for advanced detection.
\newblock {\em IEEE Transactions on Applied Superconductivity}, 32(4):1--5, June 2022.

\bibitem{OpenCV}
OpenCV.
\newblock https://docs.opencv.org/4.x/index.html.

\bibitem{scikit}
scikit learn.
\newblock https://scikit-learn.org/stable/.

\end{thebibliography}

\newpage

\appendix

\section{Additional Defect Detection Results}
\begin{figure*}[!h]
\centering{\includegraphics[width=0.95\textwidth]{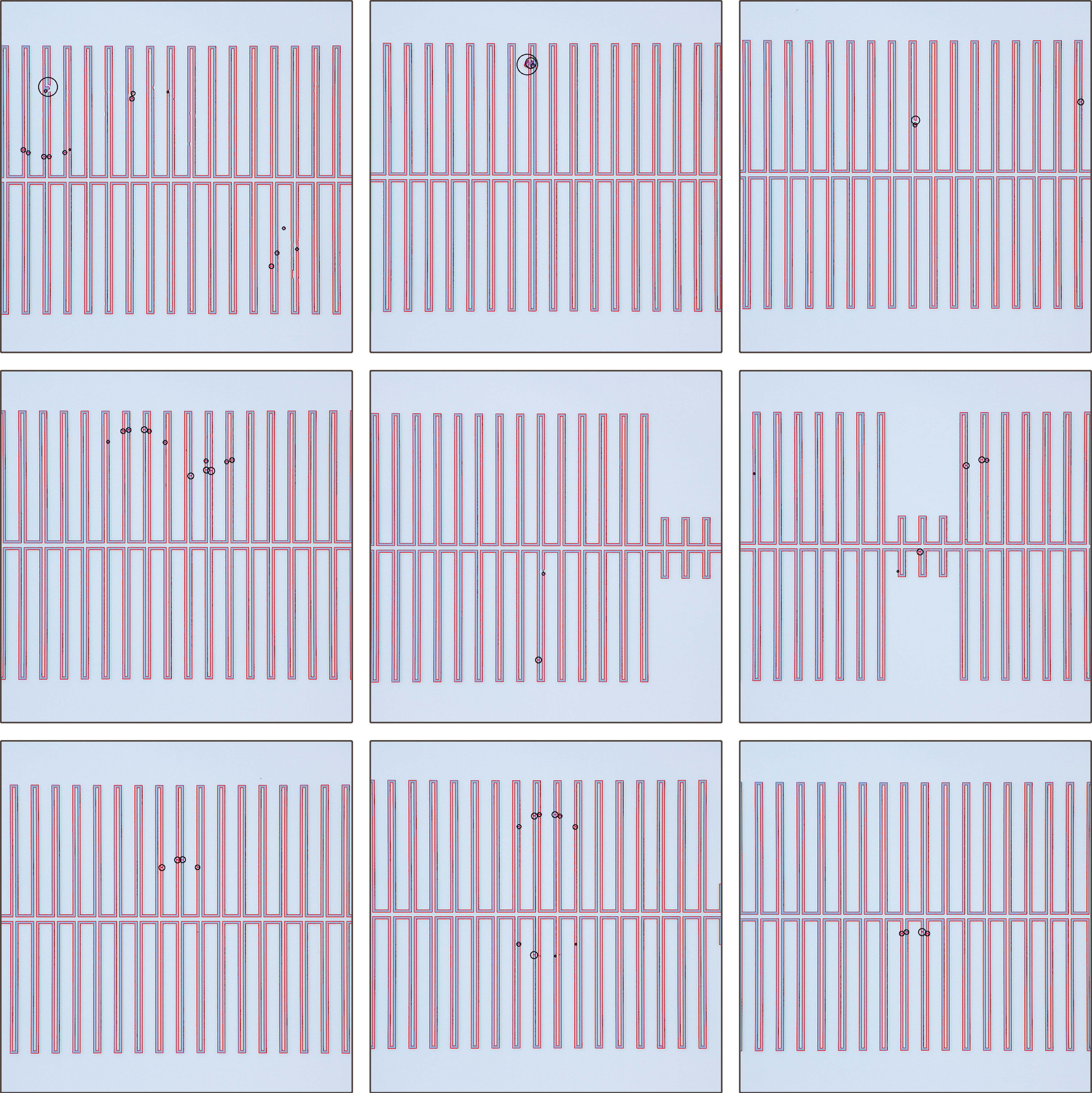}} 
\caption{Example images featuring successful defect detection using image-processing techniques on the KTWPA line, some of which are associated with nanometer-scale bridges of NbTiN defects.}
\label{fig:sup1}
\end{figure*}

To further demonstrate the capability of defect detection through image processing techniques, we present additional results in Fig.~\ref{fig:sup1}. Most detections correspond to nanometer-scale defective features that are not easily discernible through manual inspection under an optical microscope but are effectively identified by the algorithm described in section.~\ref{sec:img_proc}. The small artefacts shown in Fig.~\ref{fig:sup1}, although not exceeding the nanometer scale, still pose a potential short-circuit risk. While some of them appear to not form a full connection when imaged with the SEM, further analysis a sample defect with EDS revealed traces of the NbTiN film, suggesting potential formation of a bridge that could lead to a short circuit in KTWPA samples.

\end{document}